\documentclass[prl,twocolumn,superscriptaddress]{revtex4-1}  

\usepackage{graphicx}
\usepackage{amsmath}
\usepackage{amssymb}
\usepackage{bm}        
\usepackage{txfonts}
\usepackage{multirow}
\usepackage{mathptmx}  

\begin{document}

\title{Ultracold anions for high-precision antihydrogen experiments}

\author{G.~Cerchiari}
\affiliation{Max Planck Institute for Nuclear Physics, Saupfercheckweg~1, 69117~Heidelberg, Germany}
\author{A.~Kellerbauer}
\altaffiliation{Corresponding author}
\email{a.kellerbauer@cern.ch}
\affiliation{Max Planck Institute for Nuclear Physics, Saupfercheckweg~1, 69117~Heidelberg, Germany}
\author{M.~S. Safronova}
\affiliation{Department of Physics and Astronomy, University of Delaware, \\217 Sharp Lab, Newark, DE~19716, USA}
\affiliation{Joint Quantum Institute, National Institute of Standards and Technology and the University of Maryland, Gaithersburg, MD~20742, USA}
\author{U.~I. Safronova}
\affiliation{Physics Department, University of Nevada, Reno, NV~89557, USA}
\author{P.~Yzombard}
\affiliation{Max Planck Institute for Nuclear Physics, Saupfercheckweg~1, 69117~Heidelberg, Germany}

\date{\today}

\begin{abstract}
Experiments with antihydrogen ($\overline{\text{H}}$) for a study of matter--antimatter symmetry and antimatter gravity require ultracold $\overline{\text{H}}$ to reach ultimate precision. A promising path towards anti-atoms much colder than a few kelvin involves the pre-cooling of antiprotons by laser-cooled anions. Due to the weak binding of the valence electron in anions -- dominated by polarization and correlation effects -- only few candidate systems with suitable transitions exist. We report on a combination of experimental and theoretical studies to fully determine the relevant binding energies, transition rates and branching ratios of the most promising candidate La$^{-}$. Using combined transverse and collinear laser spectroscopy, we determined the resonant frequency of the laser cooling transition to be $\nu = 96.592\,713(91)$~THz and its transition rate to be $A = 4.90(50) \times 10^{4}$~s$^{-1}$. Using a novel high-precision theoretical treatment of La$^-$ we calculated yet unmeasured energy levels, transition rates, branching ratios, and lifetimes to complement experimental information on the laser cooling cycle of La$^-$. The new data establish the suitability of La$^-$ for laser cooling and show that the cooling transition is significantly stronger than suggested by a previous theoretical study.
\end{abstract}

\maketitle


Antihydrogen ($\overline{\text{H}}$), the simplest antimatter atom, is an ideal laboratory to search for deviations from the symmetry between matter and antimatter (CPT invariance) and the gravitational acceleration of antimatter (Weak Equivalence Principle -- WEP). Since the first production of cold $\overline{\text{H}}$ in 2002 \cite{bib:amor2002}, several follow-up experiments at CERN's Antiproton Decelerator (AD) are aiming to study $\overline{\text{H}}$ by laser or microwave spectroscopy (ALPHA \cite{bib:cesa2009}, ATRAP \cite{bib:gabr2008} and ASACUSA \cite{bib:kuro2014}) and gravimetry (AEGIS \cite{bib:kell2008, bib:dose2012} and GBAR \cite{bib:inde2014}). The production of large amounts of $\overline{\text{H}}$, its confinement \cite{bib:andr2010}, the observation of the $\overline{\text{H}}$ hyperfine spectrum \cite{bib:ahma2017b}, as well as 1S--2S laser spectroscopy \cite{bib:ahma2017a}, are important milestones towards these physics goals.

Presently $\overline{\text{H}}$ is produced at best at the temperature of the traps confining its constituents, antiprotons ($\bar{p}$) and positrons ($e^{+}$). With cryogenic traps cooled by liquid helium, the $\overline{\text{H}}$ temperature is therefore limited to $\approx 4$--$10$~K. Trapping in a shallow magnetic trap can further reduce the temperature to $\approx 0.7$~K. Several techniques for the production of significantly colder $\overline{\text{H}}$ have been proposed: (1) Direct laser cooling of $\overline{\text{H}}$ with a Lyman-$\alpha$ laser \cite{bib:wu__2011,bib:donn2013}; (2) Sympathetic cooling of {$\overline{\text{H}}^+$} with laser-cooled atomic \textit{cations}, followed by photodetachment \cite{bib:hili2014}; and (3) Production of $\overline{\text{H}}$ from $\bar{p}$ sympathetically cooled by laser-cooled \textit{anions} \cite{bib:kell2006, bib:yzom2015}.

The latter technique is particularly suited for the AEGIS experiment, where $\overline{\text{H}}$ will be created by resonant charge exchange of positronium (Ps) and pre-cooled $\bar{p}$. Due to the large $\bar{p}$--$e^{+}$ mass ratio, the final $\overline{\text{H}}$ temperature is near the initial $\bar{p}$ temperature. This scheme is contingent on the availability of a fast electric-dipole transition in a negative ion. Allowed electronic transitions are rare in atomic anions, where the (weak) binding of the valence electron is dominated by polarization and correlation effects. We initially studied Os$^-$, the first anion system in which an electric-dipole transition was identified \cite{bib:bilo2000}, but ultimately discarded it as a potential laser cooling candidate due to the low transition rate of $\approx 50$~Hz \cite{bib:warr2009, bib:fisc2010, bib:kell2011, bib:kell2014}.

More recently, a new candidate transition connecting the $5d^2\,6s^2$ $^3$F$_2^e$ ground and the $5d\,6s^2\,6p$ $^3$D$_1^o$ excited state in La$^-$ (at $\lambda \approx 3.1$~$\mu$m) was predicted \cite{OmaBec10} and observed \cite{WalGibMat14}. Our transition frequency measurement by collinear laser spectroscopy \cite{bib:jord2015} confirmed the initial observation, but the lack of a cross-section measurement left the crucial question of the cooling rate unanswered. In this article we report on a new experiment, which makes use of \textit{transverse} spectroscopy to directly measure the resonant cross-section, as well as new high-precision theoretical calculations that provide additional important information on the laser cooling cycle. Anion laser cooling holds the potential to allow the production of ultracold ensembles of any negatively charged species, opening a new frontier of ultracold science and enabling important fundamental-physics tests.

\begin{figure}
\includegraphics[width=82mm]{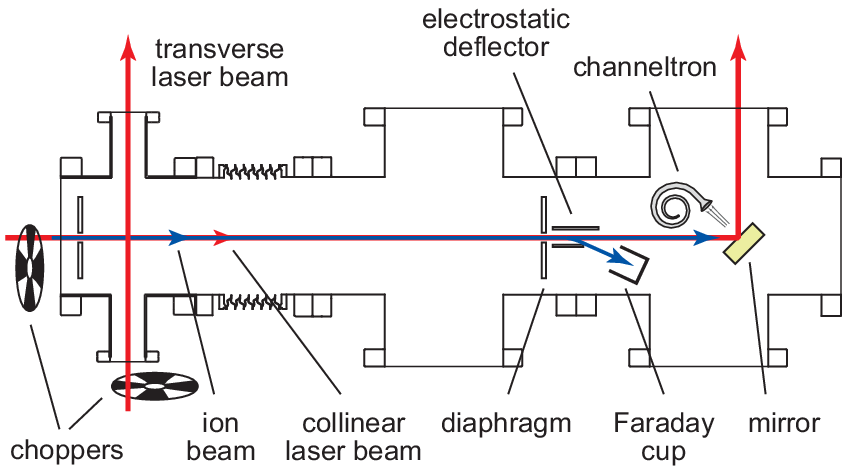}
\caption{(Color online.) Sketch of the collinear/transverse laser spectroscopy setup (top view) with an overlap region of 730~mm. The ion beam is indicated by a blue line, the laser beams by red lines.\label{fig:setup}}
\end{figure}

The cross-section of the $^3\text{F}_2^{e}\ \longleftrightarrow\ ^3\text{D}_1^{o}$ transition was measured with a setup similar to that described in Ref.~\cite{bib:jord2015}, but improved in several important aspects. The main difference is the addition of transverse laser excitation (see Fig.~\ref{fig:setup}), a prerequisite for a precise cross-section measurement.
A continuous ion beam from a Cs sputter source accelerated to 7~keV is collimated by two circular apertures with radius $r = 3.25$~mm placed a distance $d = 730$~mm apart. After the last aperture two parallel-plate electrodes deflect ions into a Faraday cup for current monitoring. Neutral particles pass straight ahead and are counted by a channeltron detector via secondary-electron emission from the surface of a gold mirror. The mirror is inclined by 45$^{\circ}$ with respect to the incoming beam direction. Two 25.4~mm windows are used to couple light collinearly with the ion beam path via a gold mirror. A second pair of windows allows orthogonal access 60~mm downstream of the first collimating aperture.

The laser light for both collinear and transverse illumination of the ion beam is produced by a continuous-wave optical parametric oscillator system (Aculight Argos 2400) able to generate $1.5$~W of light near the wavelength of interest with a nominal linewidth $< 1$~MHz. The radial power profile of the transverse laser beam was measured to be Gaussian with a width $2 \sigma_{r} = 2.69(5)$~mm. The tranverse intensity is set by a half-wave plate and polarizer. The intensity of the collinear beam is not modified. The beam is, however, reflected once along its path.

In all experiments, the laser frequency was set to (or scanned around) the $^3F_2^{e} \longleftrightarrow {^3D_1^{o}}$ transition. After resonant excitation, the absorption of a second photon of energy $\approx 400$~meV leads to photodetachment. Due to the larger interaction volume and time, the detachment step is much more likely to be caused by the collinear beam. Also, the collinear beam is out of resonance with the transition due to the Doppler effect. Consequently, independence of the two excitation steps was assumed. When the transverse laser frequency matches the transition frequency, the neutral count rate on the channeltron increases.

Both laser components are chopped, at 25~Hz (transverse) and 5~Hz (collinear), and the integrated neutral counts are recorded every 2~s. The final signal is obtained by subtracting the counts with either laser illuminating the ions from those with both lasers on. To account for the background, the counts with both lasers off are once again added. As the collinear laser beam, the transverse beam can also be reflected back onto itself. A shutter with a 4~s period obstructs the reflected beam, allowing both single and double transverse illumination.


A prerequisite for the cross-section measurement was the confirmation of the transition frequency of our previous work \cite{bib:jord2015}. The alignment of the primary and reflected transverse laser beams was optimized by comparing the single- and double-pass data. This eliminates any frequency shift due to non-perpendicular alignment of the ion and laser beams and allows a reduction of the observed linewidth by saturation spectroscopy. With single-pass transverse spectroscopy, the width of individual resonances reached $\Gamma_{\text{res}} > 150$~MHz (FWHM of the Gaussian) due to power broadening. In double-pass configuration, the width was reduced to 64.3(1.5)~MHz (FWHM of the Lorentzian) in the Doppler-free spectrum. This value is limited by the interaction time of the ions with the transverse laser (time-of-flight broadening), but improves upon our previous collinear-spectroscopy result of $\Gamma_{\text{res}} \approx 75$~MHz.

\begin{figure}
\includegraphics[width=80mm]{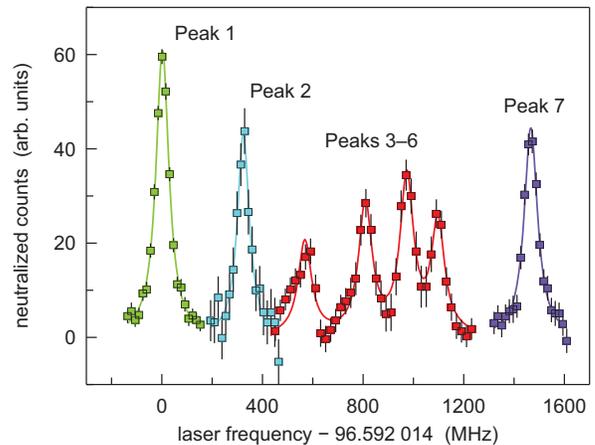}
\caption{(Color online.) Composite graph of the $^3\text{F}_2^{e} \longrightarrow {^3\text{D}_1^{o}}$ resonance in La$^-$, resolving its hyperfine structure. Different colors correspond to four different acquisitions combined in this plot. Solid lines are Lorentzian fits of the peaks.}
\label{hyperfine_resonance}
\end{figure}

Figure~\ref{hyperfine_resonance} shows the hyperfine spectrum of the $^3\text{F}_2^e\ \longrightarrow\ ^3\text{D}_1^o$ transition, assembled from separate measurements of individual hyperfine transitions or groups thereof.
The measured frequencies (see Tab.~\ref{transition_frequencies}) agree well with our previous values from collinear spectroscopy \cite{bib:jord2015}.%
\begin{table}
 \vspace{-\abovecaptionskip}
 \caption{Hyperfine transition frequencies relative to $11/2 \longleftrightarrow 9/2$ [at $\nu = 96.592\,014(05)(75)$~THz]. Transitions are listed by total angular momentum $F$ of the ground (left) and excited states (right).}
 \label{transition_frequencies}
\vspace{\abovecaptionskip}
 \centering
 \begin{tabular*}{\linewidth}{@{\extracolsep{\fill}}l l c c}
 \hline \hline
 \multicolumn{1}{c}{\rule{0pt}{2.5ex}Peak} & \multicolumn{1}{c}{Hyperfine} & \multicolumn{2}{c}{Transition frequency (MHz)}\\
 \multicolumn{1}{c}{no.} & \multicolumn{1}{c}{transition} & Exp.~\cite{bib:jord2015} & this work\\[0.5ex]
 \hline
 \rule{0pt}{2.5ex}1 & $11/2 \longleftrightarrow 9/2$ & 0.0(5.8) & 0.0(3.5)\\
 2 & $9/2 \longleftrightarrow 7/2$ & 324.8(5.8) & 321.6(4.6)\\
 3 & $7/2 \longleftrightarrow 5/2$ & 604.1(5.9) & 585.8(5.8)\\
 4 & $9/2 \longleftrightarrow 9/2$ & 825.1(5.8) & 814.2(5.2)\\
 5 & $7/2 \longleftrightarrow 7/2$ & 990.1(5.9) & 977.2(5.2)\\
 6 & $5/2 \longleftrightarrow 5/2$ & 1116.2(6.1) & 1102.5(5.2)\\
 7a & \multirow{3}{*}{$\left.\hspace{-\nulldelimiterspace}\begin{tabular}{@{}l@{}l@{}c@{}c@{}}$3/2 \longleftrightarrow 5/2$ \\ $5/2 \longleftrightarrow 7/2$ \\$7/2 \longleftrightarrow 9/2$\end{tabular}\hspace{1em}\right\}$} & & \\
 7b & & 1480.2(5.8) & 1467.6(4.0)\\
 \rule[-1.2ex]{0pt}{0ex}7c & & \\
 \hline \hline
\end{tabular*}
\end{table}
As in the prior work, the internal structure of the highest-frequency peak (consisting of three hyperfine transitions) was not resolved. The center-of-gravity frequency was found to be 96.592\,713(52)(75)~THz, where the first number in parentheses represents the statistical uncertainty and the second the systematic uncertainty, in good agreement with our prior result of 96.592\,80(10)~THz.

\begin{figure}
\includegraphics[width=80mm]{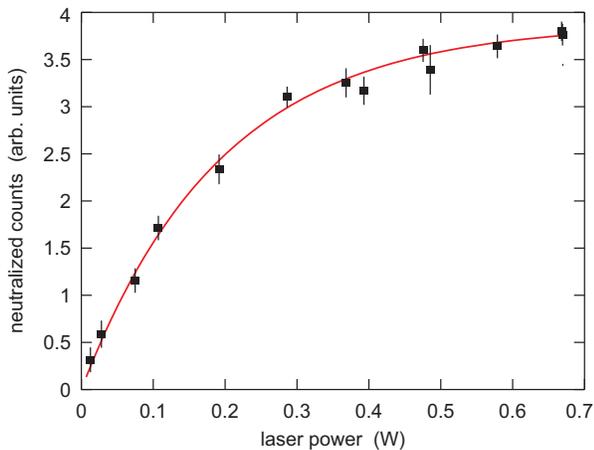}
\caption{(Color online.) Neutral atoms detected as a function of transverse laser beam power. The solid red line is a fit with the analytical rate equation solution $N(W_0)$, as described in the text.}
\label{fluence_scan}
\end{figure}

The resonant cross-section $\sigma$ was measured using the hyperfine transition with angular momentum quantum numbers $F = 11/2 \longleftrightarrow 9/2$ (Peak~1). In this measurement, the transverse reflected laser was blocked. The relative change of neutrals $N$ was monitored as a function of the laser power $W_0$, as shown in Fig.~\ref{fluence_scan}.%
The relation linking those quantities was obtained by analytically solving the associated rate equations, see Supplemental material. Several $N\left(W_0\right)$ graphs were recorded for slightly different spatial overlap between the ion and transverse laser beams by varying the vertical laser incidence angle.

Care was also taken to take into account possible fluctuations in ion current. The dependency $N\left(W_0\right)$ was recorded by scanning the curve from high to low power and back while ensuring the consistency of data at the beginning and at the end of the sequence. Finally, an observed cross-section of $\sigma = 1.0(1) \times 10^{-12}$~cm$^2$ was obtained by interpolating the data to the maximal overlap. With an experimental transition width $\Gamma_{\text{res}} = 119.6(2.5)$~MHz (without power broadening), we find an excited-state lifetime of $\tau_\text{exp} = c^{2}/(4 \pi^{2} \sigma \nu^{2} \Gamma_\text{res}) = 20.4(2.1)$~${\mu}$s.


The complicated electronic structure of the lanthanides (with large electronic correlations) presents major obstacles to calculating La$^-$ properties. While the best prior calculation of bound states by the relativistic configuration interaction (RCI) method \cite{OMaBec09} helped identify measured lines of La$^-$ spectra \cite{WalGibMat14}, differences between theoretical and measured excitation energies were as high as 33\%. The guiding principle of our approach is to treat La$^-$ as a Xe-like core with 54 electrons and four additional valence electrons. Thus, the treatment of electronic correlations separates into two problems: (1) Strong valence--valence correlations and (2) Inclusion of core--valence correlations for such a large core. The main reason for the previous discrepancy with experiment is the omission of the latter, whereas we use a hybrid approach that efficiently treats these two correlations by separate methods \cite{SafKozJoh09}.

The first problem is treated by a very-large-scale CI method in the four-electron valence space. The many-electron wavefunction is obtained as a linear combination of all distinct four-electron states $\Phi_i$ of angular momentum $J$ and parity.
%
%
In the usual implementation of the CI, the energies and wavefunctions of the low-lying states are determined by diagonalizing the Hamiltonian $H = H_{1} + H_2$, where $H_{1}$ is the one-body part of the Hamiltonian and $H_2$ the two-body part, which contains Coulomb and Breit matrix elements. We replace this bare Hamiltonian by the effective one, $H_{1,2}^\text{eff} \rightarrow H_{1,2} + \Sigma_{1,2}$,
%
%
where the $\Sigma$ corrections incorporate single and double excitations from \textit{all} core shells to all basis set orbitals, efficiently solving the second problem. In this way, the properties of a lanthanide anion can be calculated to high precision for the first time: the accuracy of our theoretical energies is 0.2--2\%. Moreover, our method yields uncertainty estimates of yet unmeasured values of other energies and transition properties.

Such an effective Hamiltonian $H^\text{eff}$ can be constructed either using second-order perturbation theory (CI+MBPT \cite{DzuFlaKoz96,KozPorSaf15}) or by the more accurate all-order coupled-cluster method (CI+all-order \cite{SafKozJoh09}). The CI method is applied as usual with the modified $H^\text{eff}$ to obtain improved energies and wavefunctions. Other properties, such as transition matrix elements, can be determined using the resulting wavefunctions. While the CI+all-order method has been applied to neutral atoms and ions with few valence electrons \cite{SafSafCla14,SafSafCla15,PorKozSaf16}, it has never been considered for negative ions, as such computations for weakly bound states were generally assumed to be numerically unstable.

The weak binding of the valence electron leads to poor convergence of the valence CI as additional configurations are included, yielding spurious low-lying configurations in bound spectra. The number of required configurations $\Phi_i$ grows exponentially with the number of valence electrons. We developed an algorithm to efficiently select dominant configurations and performed extensive tests of our method. To ensure the completeness of the CI space, we carried out large-scale CI+all-order calculations with an increasing size of the four-electron configuration space to ensure that it was sufficiently large. We also performed CI+MBPT calculations to evaluate the importance of higher-order corrections to the effective Hamiltonian and evaluate the uncertainties of the results.


The calculation of electron affinities is very sensitive to the details of the computations of the neutral and negative-ion energies. In our method, CI+all-order calculations for La and La$^-$ share the same effective Hamiltonian constructed in the Dirac--Fock potential of the same Xe core. Therefore, they have the same core energy, and the affinity is calculated as the difference of the ground-state valence energies of La and La$^-$. Our affinity value $EA = 560(14)$~meV disagrees with the experimental result of 470(20)~meV \cite{CovCalTho98} (but agrees with the theoretical work of Ref.~\cite{OMaBec09} and the experimental results of Ref.~\cite{WalGibMat14}), so we carried out a detailed evaluation of the accuracy of our value, as described in Supplemental material.

Our CI+all-order results on bound energy levels are given in Tab.~\ref{th_tab1}.
\begin{table}
\centering
\vspace{-\abovecaptionskip}
\caption{\label{th_tab1}Level energies in La$^-$ evaluated using the CI+all-order method, relative to the ground state. The last column gives the difference with experiment.}
\vspace{\abovecaptionskip}
\begin{tabular*}{\linewidth}{@{\extracolsep{\fill}} l@{~}lcccc }
\hline \hline
\multicolumn{2}{c}{\rule{0pt}{2.5ex}Config.} & \multicolumn{3}{c}{Level energy (meV)} & \\
 & & Theo.~\cite{OMaBec09} & Exp.~\cite{WalGibMat14} & this work & Diff.\\[0.5ex]
\hline
\rule{0pt}{2.5ex}$6s^25d^2$ & $^3F_{3}$ & 67 & 83.94 & 83.75 & $+0.2$\%\\
 & $^3F_{4}$ & 135 & 172.9 & 174.8 & $-1.1$\%\\
 & $^1D_{2}$ & 286 & & 328.1 & \\
 & $^3P_{0}$ & 417 & & 410.2 & \\
 & $^3P_{1}$ & 442 & & 440.4 & \\
 \rule[-1.2ex]{0pt}{0ex} & $^3P_{2}$ & 493 & & 504.7 & \\
\hline
\rule{0pt}{2.5ex}$6s^25d6p$ & $^1D_{2}$ & 111 & & 217.9 & \\
 & $^3F_{2}$ & 259 & 343.7 & 345.8 & $-0.6$\%\\
 & $^3F_{3}$ & 305 & 383.9 & 389.1 & $-1.4$\%\\
 & $^3D_{1}$ & 337 & 399.4 & 406.9 & $-1.9$\%\\
 & $^3D_{2}$ & 396 & 470.6 & 478.5 & $-1.7$\%\\
 & $^3F_{4}$ & 406 & 496.2 & 502.9 & $-1.4$\%\\
 & $^3P_{0}$ & 535 &       & 548.6 & \\
 \rule[-1.2ex]{0pt}{0ex} & $^3D_{3}$ & 461 & 538.8 & 549.3 & $-2.0$\%\\
\hline \hline
\end{tabular*}
\end{table}
The values are in excellent agreement with experimental data. We predict a number of yet unmeasured La$^-$ bound states, in particular the $\text{6s}^2\text{5d}^2\,\,^3P_J$ triplet. Our calculations show that the $^3P^e_0$ level is much closer to the upper cooling transition level $^3D^o_1$ than indicated by Ref.~\cite{OMaBec09}. The energy difference is on the order of our computational accuracy. The electric-dipole (E1) matrix element for the $^3P_0^{e} \longleftrightarrow {^3D_1^{o}}$ transition is 1.17~$e\,a_0$, where $a_0$ is the Bohr radius. If the metastable $^3P_0^e$ level is energetically lower, the branching ratio to that state strongly depends on the transition energy. We estimate the branching to be very small, $3\times10^{-7}$ to $1\times10^{-6}$ (for a separation energy of 3--5~meV).


We calculated the E1 matrix elements for allowed La$^-$ transitions using both CI+MBPT and three variants of CI+all-order wavefunctions with increasing numbers of configurations. The relevant transition rates, branching ratios and lifetimes are given in Tab.~\ref{th_tab2}.
\begin{table}
\centering
\vspace{-\abovecaptionskip}
\caption{\label{th_tab2}Transition rates $A_r$, branching ratios, and lifetimes $\tau$ of transitions in La$^-$ calculated using the CI+all method. Numbers in brackets represent powers of 10.}
\vspace{\abovecaptionskip}
\begin{tabular*}{\linewidth}{@{\extracolsep{\fill}}l@{}ll@{}lccc}
 \hline \hline
 \multicolumn{2}{c}{\rule{0pt}{2.5ex}Upper} & \multicolumn{2}{c}{Lower} & \multicolumn{1}{c}{$A_{r}$} &  \multicolumn{1}{c}{Branching} & \multicolumn{1}{c}{$\tau$}\\
 \multicolumn{2}{c}{level} & \multicolumn{2}{c}{level} & \multicolumn{1}{c}{(s$^{-1}$)} & \multicolumn{1}{c}{ratio} & \multicolumn{1}{c}{}\\[0.5ex]
 \hline
 \rule{0pt}{2.5ex}$6s^25d6p$ & $^3D_{1}$ &  $6s^25d^2$ & $^3F_{2}$ & 4.54[$+4$] & 0.999974 & 22.1~$\mu$s\\
 \rule[-1.2ex]{0pt}{0ex} & & $6s^25d^2$ & $^1D_{2}$ & 1.18[$\mspace{12mu}0$] & 0.000026 & \\
 \hline
 \rule{0pt}{2.5ex}$6s^25d^2$ & $^1D_{2}$ & $6s^25d6p$ & $^1D_{2}$ & 1.95[$\mspace{12mu}0$] & 0.956  & 489~ms\\
 \rule[-1.2ex]{0pt}{0ex} & & $6s^25d^2$ & $^3F_{2}$ & 9.00[$-2$] & 0.044 & \\
 \hline
 \rule{0pt}{2.5ex}$6s^25d6p$ & $^1D_{2}$ & $6s^25d^2$ & $^3F_{2}$ & 1.68[$+2$] & 0.791 & 4.71~ms\\
 \rule[-1.2ex]{0pt}{0ex} & & $6s^25d^2$ & $^3F_{3}$ & 4.44[$+1$] & 0.209 & \\
 \hline
 \rule[-1.2ex]{0pt}{0ex}\rule{0pt}{2.5ex}$6s^25d^2$ & $^3F_{3}$ & $6s^25d^2$ & $^3F_{2}$ & 7.56[$-3$] & 1.0 & 132~s\\
 \hline
 \rule{0pt}{2.5ex}$6s^25d6p$ & $^3D_{2}$ & $6s^25d^2$ & $^3F_{2}$ & 4.50[$+3$] & 0.1059 & 23.5~$\mu$s\\
  & & $6s^25d^2$ & $^3F_{3}$ & 3.79[$+4$] & 0.8924 & \\
  & & $6s^25d^2$ & $^1D_{2}$ & 4.41[$+1$] & 0.0010 & \\
 \rule[-1.2ex]{0pt}{0ex} & & $6s^25d^2$ & $^3P_{1}$ & 2.75[$+1$] & 0.0006 & \\
 \hline \hline
 \end{tabular*}
\end{table}
We used experimental energies from Ref.~\cite{WalGibMat14} where available. The uncertainties were determined as described in Supplemental material. Ultimately, we estimate the uncertainty of the laser cooling transition rate to be 4\%. The corresponding lifetime $\tau_{\text{theo}} = 22.1(9)$~$\mu$s can be directly compared to the experimental result $\tau_{\text{exp}} = 20.4(2.1)$~$\mu$s because in the absence of other allowed decays, the (measured) $11/2 \longrightarrow 9/2$ transition rate is equal to the total decay rate from the excited state. Within their respective uncertainties, our experimental and theoretical values are in excellent agreement.


The determined transition rates and branching ratios are crucial for the implementation of the cooling cycle. Our value for the laser cooling transition rate is $A_r \approx 4.5$--$4.9 \times 10^4$~s$^{-1}$, about 50\% larger than the previous prediction \cite{OmaBec10}. The laser-cooled ensemble will thus reach the Doppler temperature $T_D = 0.17$~$\mu$K (or the equilibrium temperature in case of competing heating processes) faster than previously assumed. For instance, one-dimensional cooling of an ensemble of La$^{-}$ ions from 100~K to $T_D$ will require the absorption of $8.4 \times 10^{4}$ photons, and hence take $3.7$~s in saturation.

\begin{figure}
\includegraphics[width=80mm]{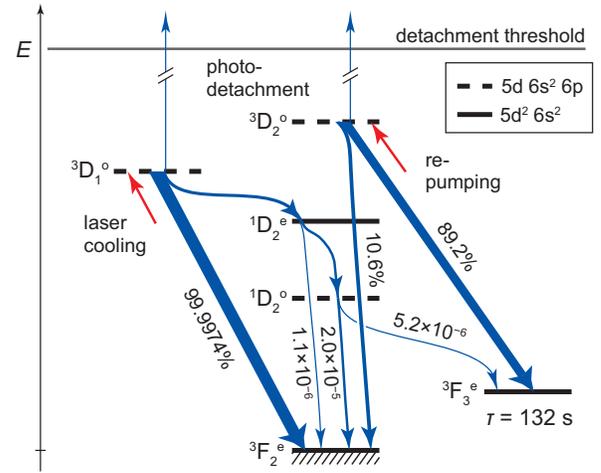}
\caption{Partial energy level diagram of La$^-$ (energies to scale). The relevant decay branches from the $^3$D$_1^o$ excited state of the laser cooling transition, as well as from the $^3$D$_2^o$ excited state of the repumping transition, are indicated. Thicknesses of blue arrows are indicative of branching ratios, but not to scale.\label{fig:branchings}}
\end{figure}

Another important question is the degree to which the cooling cycle is closed. We can fully deduce all relevant branchings from our present results, as illustrated in the partial energy level diagram of Fig.~\ref{fig:branchings}.%
(Branchings at least two orders of magnitude smaller than those indicated have been neglected.) The excited $^3\text{D}_1^{o}$ state will decay to the $^3\text{F}_2^{e}$ ground state in all but $2.6 \times 10^{-5}$ of cases. However, only a total $5.2 \times 10^{-6}$ branching to the metastable $^3\text{F}_3^{e}$ state ($\tau = 132$~s) is potentially problematic. After scattering $8.4 \times 10^{4}$ photons, roughly 40\% of all anions end up in the metastable state. If necessary, the $^3\text{F}_3^{e}\ \longrightarrow {^3\text{D}_2^{o}}$ transition can be repumped, as shown in the figure.

Finally, photodetachment from an excited state may remove La$^{-}$ ions from the trap. The photodetachment cross-section from the $^3$D$_1^o$ or the $^3$D$_2^o$ state is about five orders of magnitude smaller than that of resonant excitation. Its rate also depends on the fraction of time during which ions populate excited states, and hence on the saturation of the cooling transition. Assuming a saturation parameter $s = 2$, the photodetachment rate is $0.024$~s$^{-1}$ and the branching to photodetachment $5.3 \times 10^{-7}$. Since at most a few repumping cycles will be required, photodetachment from the $^3$D$_2^o$ state is negligible. Hence, a fraction of about 10\% of La$^{-}$ ions will be lost by the time the remaining ensemble has been cooled.

\begin{acknowledgments}
We thank the MPIK accelerator group and workshop for invaluable assistance with the ion source and Sergey Porsev and Mikhail Kozlov for helpful discussions. Experimental work was supported by the European Research Council (ERC) under Grant No. 259209 (UNIC). Theoretical work was performed under the sponsorship of the U.S. Department of Commerce, National Institute of Standards and Technology.
\end{acknowledgments}

\end{document}